\documentclass[lettersize,journal]{IEEEtran}
\usepackage{amsmath,amsfonts}
\usepackage{algorithmic}
\usepackage{array}
\usepackage[caption=false,font=normalsize,labelfont=sf,textfont=sf]{subfig}
\usepackage{textcomp}
\usepackage{stfloats}
\usepackage{url}
\usepackage{verbatim}
\usepackage{graphicx}
\def\BibTeX{{\rm B\kern-.05em{\sc i\kern-.025em b}\kern-.08em
    T\kern-.1667em\lower.7ex\hbox{E}\kern-.125emX}}
\usepackage{balance}

\usepackage{xcolor}
\usepackage{soul}
\usepackage{algorithmic}
\usepackage[ruled]{algorithm2e} 

\usepackage{booktabs}
\usepackage{multirow}

\begin{document}
\title{Statistical Hardware Design With Multi-model Active Learning}
\author{Alireza Ghaffari, Masoud Asgharian and Yvon Savaria.
\thanks{Alireza Ghaffari and Yvon Savaria are with the department of Electrical Engineering Polytechnique Montreal, and Masoud Asgharian is with the department of Mathematics and Statistics, McGill Univesrity. \textit{Corresponding Author: seyed-alireza.ghaffari@polymtl.ca}.}}

\markboth{}
{How to Use the IEEEtran \LaTeX \ Templates}

\maketitle

\begin{abstract}
With the rising complexity of numerous novel applications that serve our modern society comes the strong need to design efficient computing platforms. Designing efficient hardware is, however, a complex multi-objective problem that deals with multiple parameters and their interactions. Given that there are a large number of parameters and objectives involved in hardware design, synthesizing all possible combinations is not a feasible method to find the optimal solution. One promising approach to tackle this problem is statistical modeling of a desired hardware performance. Here, we propose a model-based active learning approach to solve this problem. Our proposed method uses Bayesian models to characterize various aspects of hardware performance. We also use transfer learning and Gaussian regression bootstrapping techniques in conjunction with active learning to create more accurate models. Our proposed statistical modeling method provides hardware models that are sufficiently accurate to perform design space exploration as well as performance prediction simultaneously. We use our proposed method to perform design space exploration and performance prediction for various hardware setups, such as micro-architecture design and OpenCL kernels for FPGA targets. Our experiments show that the number of samples required to create performance models significantly reduces while maintaining the predictive power of our proposed statistical models. For instance, in our performance prediction setting, the proposed method needs 65\% fewer samples to create the model, and in the design space exploration setting, our proposed method can find the best parameter  settings by exploring less than 50 samples.
\end{abstract}


\section{Introduction}
\IEEEPARstart{E}{fficient} hardware design is essential for various industrial and scientific applications. With the emergence of new applications such as machine learning, the Internet of Things (IoT), and 5G wireless communication, there is a strong need to design computing platforms that provide the required computational power. Designing efficient hardware is challenging and requires considering various hardware and software parameters and their interactions. In most cases, designing efficient digital hardware is a multi-objective optimization problem where the designers try to optimize multiple competing objectives, such as energy consumption and throughput. Moreover, performing a brute-force search to find the optimum design is not feasible, because synthesizing complex hardware designs is very time-consuming. In addition, designers must test a large number of parameter combinations to find the optimum solution.

Another prominent challenge of efficient hardware design is that the objective function is not smooth, and most parameters are discrete. For example, the memory delay, the number of floating-point arithmetic units, and clock frequencies are parameters that need to be tuned for throughput. Since these parameters are discrete, the objective function is not differentiable. Thus using gradient-based methods such as Stochastic Gradient Descent (SGD) to find the optimum parameter set is not possible.

For years, meta-heuristic methods, such as Genetic Algorithm, were used to perform design space exploration \cite{deb2002fast,palesi2002multi}. Although these meta-heuristic algorithms can find the optimum setup in various settings, they cannot provide a model that predicts the performance. Moreover, the heuristic methods cannot provide an abstract model to study the interactions and inter-dependencies between the parameters and their effect on the desired performance metrics. Researchers proposed Bayesian models for design space exploration tasks to solve this issue. Bayesian optimization is not only capable of finding the optimum parameter setup but is also able to reveal the interactions and inter-dependencies between the parameters of the design.  

Although research on Bayesian models has produced remarkable results for design space exploration as well as abstract modeling of parameter inter-dependencies, they incorporate two fundamental shortcomings. First, there is a need for a feedback mechanism from a real-world application to update the Bayesian model to improve accuracy. Second, Bayesian models are application specific, meaning that a model trained for a particular task cannot be used for another. Thus, there is a need to propose a \textit{transfer learning} strategy to use currently trained models to predict the performance of applications that were not previously exposed to the model. Using transfer learning is crucial in hardware design since it provides the ability to perform design space exploration for emerging applications using prior knowledge acquired from previous experiences and applications.

Here we propose a novel methodology that solves the two shortcomings mentioned above. Besides the Bayesian modeling approach, we suggest using Bayesian active learning (\cite{kandasamy2015bayesian,brochu2010tutorial}) to increase our proposed statistical model accuracy. Bayesian active learning is an iterative machine learning algorithm that actively selects some informative samples to construct the Bayesian model of the objective function. The iterative nature of the Bayesian active learning method helps the Bayesian model to learn the underlying structure of the data gradually. The ultimate goal of Bayesian active learning is to provide a surrogate function that closely resembles the system's behavior with the minimum number of samples.
In our proposed methodology, once a Bayesian model is created, it can be updated using a feedback mechanism that is provided by a real-world application. For instance, a Bayesian surrogate model can be updated using a processor or computing cluster performance metrics reflecting its performance for different computation loads. Furthermore, we propose using the Bayesian transfer learning \cite{skolidis2012transfer}, where the information of one domain can be transferred to another domain to improve the accuracy of the statistical model of the target task. Transfer learning is particularly useful in hardware design in scenarios where the target application is novel and there are limited information and data on the performance of the hardware in that application. In such cases, the information of previously trained hardware models can be leveraged to infer the performance of the new application.

To further reduce the number of samples needed to make hardware performance prediction, we suggest using \textit{Gaussian regression bootstrapping}. Once we have a Bayesian model of the hardware, it is possible to re-sample the model in order to acquire new composite samples to perform other types of statistical analysis. In this paper, we used Gaussian regression bootstrapping to generate new data to perform various regression analyses of the hardware model and showed that the results based on bootstrapped samples could closely resemble the analyses done on the real data.

To summarize, this paper makes the following contributions:
\begin{itemize}
    \item We propose a \textit{model-based active learning} framework that generates statistical models of different aspects of hardware performance.
    \item We propose \textit{Bayesian transfer learning} to effectively perform hardware design space exploration using prior hardware performance information.
    \item We use a \textit{Gaussian regression bootstrapping} method to effectively estimate the behavior of the hardware from the limited samples acquired from real-world applications.
\end{itemize}
To the best of our knowledge, this is the first time that model-based active learning has been used in conjunction with Bayesian transfer learning to provide accurate statistical hardware models for design-space exploration. Furthermore, this is the first time that Bayesian regression bootstrapping is used to estimate the behavior of specific hardware better.

The rest of the paper is structured as follows. Section \ref{sec:related-work} reviews related works in the hardware design space exploration and performance prediction. We introduce our methodology in Section \ref{sec:method}. In Section \ref{sec:dse-vs-reg}, we distinguish two important settings, notably design space exploration and performance modeling. In Section \ref{sec:results}, we show that our proposed methodology is effective for both design space exploration and performance modeling.

\section{Related works}\label{sec:related-work}

Exploring various design parameters in hardware design is a ubiquitous problem. The most obvious but inefficient solution is brute-force analysis or exhaustive search. The complexity of brute-force analysis increases exponentially with the number of parameters, making this method infeasible in practical problems. For years, researchers have used meta-heuristic methods such as Genetic Algorithms (GAs) \cite{palesi2002multi} to perform design space exploration and obtain the Pareto optimal solutions. Although meta-heuristic methods effectively find optimum solutions, they do not provide a mathematical model reflecting inter-dependencies between parameters and their effects on the objectives.

Another valuable design space exploration approach is Bayesian Optimization (BO) \cite{brochu2010tutorial}. Bayesian Optimization is especially interesting for design space exploration applications since (i) it provides a Bayesian surrogate model that helps to understand the inter-dependencies of the parameters and (ii) it is an iterative method that incorporates information from previously observed samples in the surrogate model. BO has been widely used in hardware design space exploration. For instance, in \cite{mehrabi2020bayesian}, BO was found to have superior performance for design space exploration of \texttt{C/C++} High-Level Synthesis (HLS) compared to traditional search methods. Likewise, BO was used in \cite{reagen2017case} to optimize hardware accelerators for deep neural networks. The authors reported that the BO method outperforms traditional design space methods in all metrics of interest, such as accuracy and energy consumption. Furthermore, in \cite{lo2016model}, BO is deployed to tune HLS directives to achieve minimum latency.

Inspired by traditional meta-heuristic approaches, researchers have combined BO and Grey Wolf Optimization (GWO) \cite{mirjalili2014grey} to perform design space exploration for HLS of OpenCL kernels \cite{ghaffari2021efficient}. The hybrid GWO-BO method was shown to outperform both BO and traditional search methods.

Other iterative methods such as Reinforcement Learning (RL) \cite{kaelbling1996reinforcement} are also effective for design space exploration applications. An RL agent tunes hardware parameters through some iterations to maximize the reward function related to metrics of interest.  For example, in \cite{de2020automated}, the RL method is used to explore parameters that optimize the latency of deep neural networks on the ARM-Cortex-A CPUs. Likewise, in \cite{ghaffari2020cnn2gate}, a time-limited reinforcement learning \cite{mnih2016asynchronous} is used to perform design space exploration for deeply pipelined OpenCL kernels of the convolutional neural networks. In \cite{hosny2020drills}, an Advantage Actor Critic (A2C) \cite{konda2003a2c} design space exploration is proposed to minimize area subject to a timing constraint in logic synthesis. Additionally, in  \cite{kao2020confuciux}, the authors proposed an RL-based design space exploration method and used the performance predicted by their cost model to compute the reward values. Moreover, they augmented their RL algorithm with a genetic algorithm to improve the design space exploration results.

Furthermore, using other statistical methods for hardware analysis and design is commonplace in the hardware community. For instance, in \cite{lee2006accurate}, authors studied accurate regression modeling of micro-architecture energy consumption and throughput. Moreover, in \cite{liu2013learning}, Random Forest was used for HLS design space exploration. Similarly, in \cite{singh2019napel}, Random Forest regression is used for predicting instruction per cycle for near-memory computing applications. In \cite{beltrame2010decision} Markov decision process was found effective for the design space exploration of multi-processor platforms. In \cite{nia2022rethinking}, parametric bootstrapping was used to perform stochastic ordering deep learning models that run efficiently on different hardware, while the inference efficiency and training efficiency are chosen as the metrics of interest.

Our proposed method is different from the previous Bayesian methods (\cite{mehrabi2020bayesian, reagen2017case, lo2016model}) in the way that it uses Active Learning in conjunction with Bayesian Optimization to create more accurate Bayesian models. We present results showing that our method is more effective than the hybrid GWO-BO method proposed in \cite{ghaffari2021efficient}. Furthermore, we use Bayesian Transfer Learning (TL) \cite{skolidis2012transfer} and show that the performance of the parameter search is significantly improved by incorporating information from other correlated hardware design tasks. Our proposed method differs from \cite{nia2022rethinking} since it can produce Bayesian models capable of being used for \textit{Gaussian regression bootstrapping} \cite{GPRbootstrap}. Also, after performing Gaussian regression bootstrapping, it is possible to produce composite data from our Bayesian surrogate model and perform statistical regression analysis similar to (\cite{lee2006accurate, beltrame2010decision}) without requiring real-world data samples.

\section{Methodology}\label{sec:method}
Statistical modeling of hardware and related applications such as design space exploration and performance estimation can be classified as Black Box Optimization (BBO) \cite{audet2017derivative} problems. This is because the objective function and its behavior, such as smoothness, continuity, and convexity, are unknown. Thus, gradient-based methods such as Gradient Descent (GD) and Stochastic Gradient Descent (SGD) are not the best choice for these applications. On the other hand, derivative-free methods such as meta-heuristic methods and Bayesian Optimization are among the best choices for these applications.

\begin{figure*}[!t]
\centering
\includegraphics[width=0.65\linewidth]{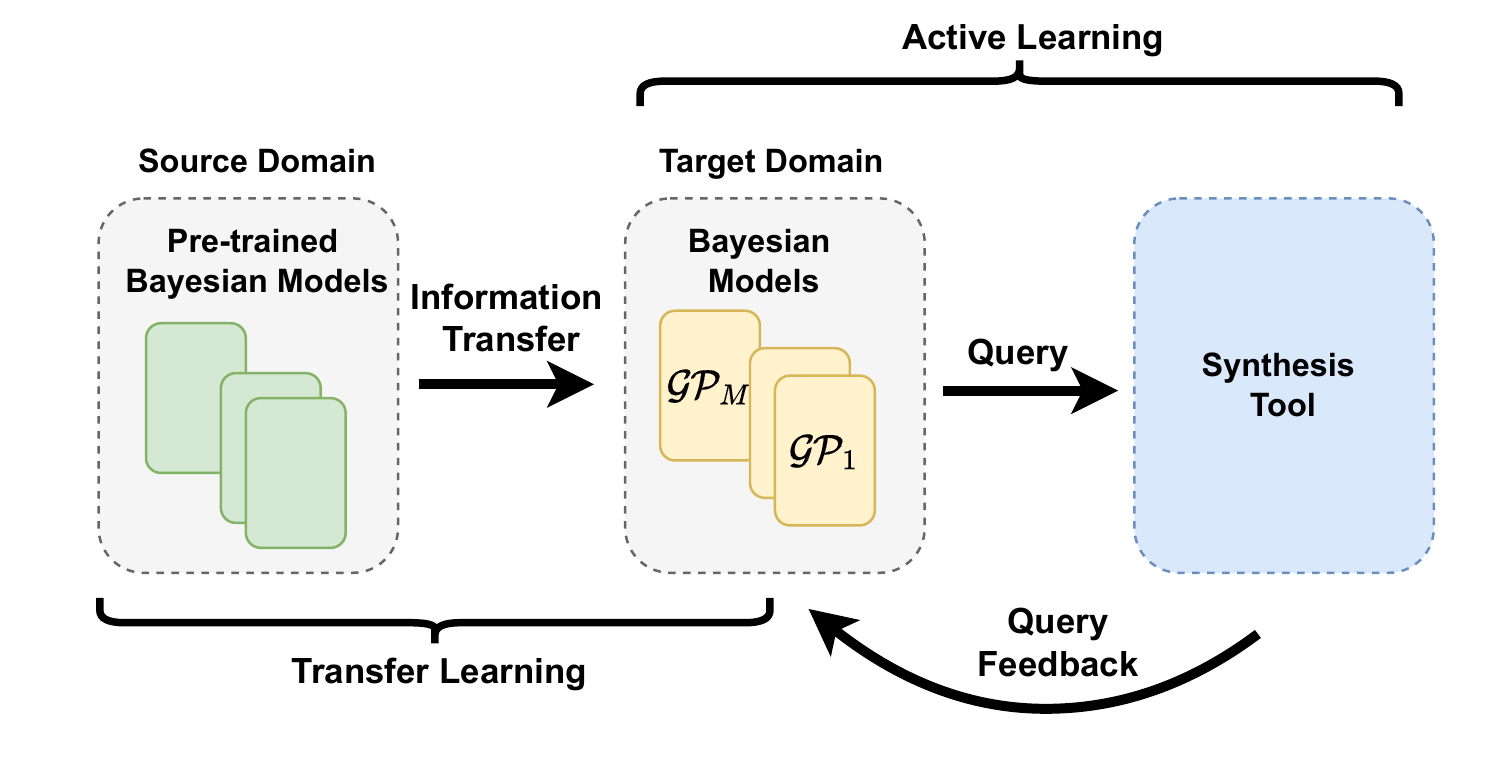}
\caption{Statistical modeling of hardware using \textit{Multi-model Bayesian Active Learning}. The framework also incorporates an optional\textit{ Bayesian Transfer Learning} setup to improve the accuracy of the Bayesian models.}
\label{fig:main-arch} 
\end{figure*}

Here we propose a \textit{Model-Based Active Learning} framework for statistical hardware  performance modeling. We further analyze this method in the context of design space exploration and performance prediction (regression analysis). In our proposed method, we use Bayesian Optimization with Gaussian Processes. We also use Bayesian Transfer Learning and Bayesian regression bootstrapping methods to further improve such models in practical applications. Fig.~\ref{fig:main-arch}  shows an overview of our proposed framework. In this figure, Bayesian models are de facto Gaussian processes that are updated iteratively using active learning. These Bayesian models can be used in design space exploration or for performance prediction. The details of our mathematical modeling are elaborated on in the following sections.

\subsection{Multi-Model Active Learning}\label{sec:methodAL}

We use active learning to update the Gaussian process models. Our proposed framework uses multiple Bayesian models for various objectives. For instance, power consumption and throughput are each associated with their corresponding models. The algorithm starts from a random set of parameters for all models. It then queries the synthesis tool and creates initial Gaussian processes. Next, the algorithm evaluates the Gaussian processes and gathers each model parameters candidate to be sent to the synthesis tool in the next query. This process continues iteratively until a particular stopping criterion is met. Furthermore, when used with active learning, the algorithm chooses the next set of parameters using the Gaussian models that are also affected by other pre-trained Gaussian processes, as explained in Section~\ref{sec:methodTL}. Also, note that our proposed algorithm considers the direction of optimization of each objective function. For instance, designers usually aim to minimize power consumption while maximizing the throughput. Therefore, our proposed algorithm chooses the sets of parameters that respect the optimization constraints of each objective function. Having multiple models associated with multiple objectives enables us to study the Pareto efficiency of the solutions. Section~\ref{sec:multi-objective-pareto} reports experimental results that validate the proposed method. Algorithm~\ref{alg:al_alg} provides a pseudo-code for our proposed algorithm. 

{
\centering
\begin{algorithm}[!b]{
\textbf{Step 1.} Randomly choose the first set of parameters $\mathbf{x}_{1:N} = \{\mathbf{x}_1, \mathbf{x}_2, \dots, \mathbf{x}_N\}$ for query. \\

\While{\textbf{stop criterion} is not satisfied} 
{
    \textbf{Step 2.} Get the query Objectives $O(\mathbf{x}_{1:N})$ . \\
    \textbf{Step 3.} For all $\{\mathcal{GP}_1, \mathcal{GP}_2, \dots \mathcal{GP}_M\}$, calculate new covariance matrix using eq~\eqref{eq:cov_matrix} \\
    \textbf{Step 4.} Update all $\{\mathcal{GP}_1, \mathcal{GP}_2, \dots \mathcal{GP}_M\}$ using eq~\eqref{eq:gp}\\
    \textbf{Step 5.} If transfer learning is needed: Update $\{\mathcal{GP}_1, \mathcal{GP}_2, \dots \mathcal{GP}_M\}$ mean vector and covariance matrix using eq~\eqref{eq:tfl}\\
    \textbf{Step 6.} Evaluate all  $\{\mathcal{GP}_1, \mathcal{GP}_2, \dots \mathcal{GP}_M\}$  and update the set $\mathbf{x}_{1:N}$
}

\textbf{return}  $\{\mathcal{GP}_1, \mathcal{GP}_2, \dots \mathcal{GP}_M\}$ 
\caption{Multi-Model Bayesian Active Learning.}
\label{alg:al_alg}
}\end{algorithm}
}

\subsection{Bayesian Optimization with Gaussian Processes}\label{sec:methodBO}

Bayesian Optimization \cite{brochu2010tutorial} is a powerful tool to find the extrema of an objective function that is hard to evaluate or does not have a close mathematical form.  Statistical modeling of hardware implementations deals with both problems simultaneously. It is well-known that synthesizing hardware models takes a very long time, and generally, there is no closed-form expression of the power consumption and latency of the resulting hardware models.

In Bayesian Optimization, the goal is to find a surrogate function $f_s$ that closely resembles the behavior of the desired objective function $O$. Note that in Bayesian Optimization, $f_s$  is a stochastic process.

Let us assume $\mathbf{x}_i$ is the \textit{i}$^{th}$ sample, and $O(\mathbf{x}_i)$ is the evaluation of the objective function for that sample. Using Bayes theorem, for a collection of $N$ samples such as $C_{1:N} = \{\mathbf{x}_{1:N}, O(\mathbf{x}_{1:N})\}$, we have

\begin{equation}
\label{eq:bayes}
P(O|C_{1:N}) \propto P(C_{1:N} | O) P(O),
\end{equation}
where $P(O|C_{1:N})$ is the posterior distribution. The posterior distribution $P$ incorporates probabilistic information about the objective function $O$. Also, the commonplace assumption in Bayesian optimization is that the objective function is continuous. This assumption does not hold in our experimental setup.  However, our goal is to perform Bayesian optimization on a discrete objective function by fitting a continuous posterior distribution $P$ that passes through the discrete points of the objective function $O$, see \cite{luong2019bayesian}.

We note that the surrogate function $f_s$ is a stochastic process. Let us further assume that $f_s$ is a Gaussian Process that provides a mean and a standard deviation at each point. i.e. 
\begin{equation}
\label{eq:gp}
f_s(\mathbf{x}) = \mathcal{GP}(\mathit{\mu}(\mathbf{x}), \mathbf{K} =\mathit{k}(\mathbf{x},\mathbf{x'}))).
\end{equation}
This assumption implies that the sampling process of $O(\mathbf{x}_i)$ is noisy with Gaussian distribution $\mathcal{N}(0,\sigma^2)$. 

In eq~\eqref{eq:gp},  $\mathcal{GP}$ denotes a Gaussian process with a mean function $\mathit{\mu}$ and a covariance operator $\mathbf{K}$.
Here,  we use the squared exponential covariance kernel $\mathit{k}$ defined below to construct the covariance matrix $\mathbf{K} $
 \begin{equation}
\label{eq:cov_matrix}
\mathbf{K}_{ij}=\mathit{k}({x}_i, {x}_j) = \mathrm{exp}{(-\frac{|{x}_i - {x}_j|^2}{2})}
\end{equation}

There are other types of covariance kernels such as \textit{Matern kernels} \cite{genton2001classes} that are suitable for non-smooth data. Although studying the effect of various covariance kernels is not the main focus of this work, we show in Appendix B that our proposed method for hardware performance modeling is robust in terms of changing the covariance kernel to Matern.

\subsection{Transfer Learning}\label{sec:methodTL}
We propose using \textit{Transfer Learning} to transfer prior information from a \textit{source} surrogate function to a \textit{target} function. Let us denote source domain $\mathcal{D}_\mathrm{source}$ that contains design parameters of our hardware design problem $\mathbf{x}=\{x_1,x_2 \dots x_p\}$. The surrogate function of source task $\mathcal{T}_\mathrm{source}$ is also denoted as $f_\mathrm{source}$. The goal is to improve the learning process of a target task $\mathcal{T}_\mathrm{source}$ with the surrogate function $f_\mathrm{target}$ and domain $\mathcal{D}_\mathrm{target}$. Note that our application considers \textit{Inductive Transfer Learning}, where $\mathcal{T}_\mathrm{source} \neq \mathcal{T}_\mathrm{target}$ and $\mathcal{D}_\mathrm{source} = \mathcal{D}_\mathrm{target}$. This means that in our experimental setup, although the tasks and their surrogate functions are different, the design parameters $\mathbf{x}=\{x_1,x_2 \dots x_p\}$ of our hardware design problems are the same.

In our proposed method, we consider \textit{Inductive Gaussian Process Transfer Learning} where we allow the mean 
function and covariance operator of the source surrogate function to affect the target surrogate function. To this end, we define
 \begin{equation}
\label{eq:tfl}
    \begin{cases}
        \mathit{\mu}^\mathrm{TL}(\mathbf{x}) = \mathit{\mu}_\mathrm{target}(\mathbf{x}) + \lambda_1 \mathit{\mu}_\mathrm{source}(\mathbf{x}) \\
        \mathbf{K}^\mathrm{TL} = \mathbf{K}^\mathrm{target} + \lambda_2\mathbf{K}^\mathrm{source}
    \end{cases}
\end{equation}
where 
$\lambda_1$,$\lambda_2$ are hyper-parameters that regularize the magnitude of the source information transferred to the target.

We emphasize that we only use transfer learning for design-space exploration experiments where finding the position of the objective function optimum point is more important than its value, see Section \ref{sec:dse-vs-reg}. This method is beneficial when the $\mathcal{T}_\mathrm{source}$ and $\mathcal{T}_\mathrm{target}$ are correlated. We will show in Section \ref{sec:dse-tl} that this assumption is true for our intended application.

\subsection{Gaussian Regression Bootstrapping}\label{sec:methodBOOTSTRAP}
In statistical hardware modeling problems, the aim is to predict the behavior of an objective function $O$. Examples of such objective functions are power consumption and latency. In practice, we have access to limited samples of this objective function because synthesizing a hardware design is a very time-consuming process. Thus, we can use the Bayesian surrogate function $f_s$ to evaluate the behavior of the objective function $O$ for new sets of parameters. This method is commonly known as \textit{Gaussian Regression Bootstrapping} \cite{GPRbootstrap} and its theoretical aspect is studied in \cite{hayashi2020random}. This is specifically beneficial  because, unlike $O$, sampling $f_s$ is inexpensive. Note that $f_s$ is a Gaussian process incorporating mean and variance information. Here, we can use the mean function $\mathit{\mu}$ of $f_s$ for resampling the objective function $O$ 

\begin{equation}
\label{eq:gp-bootstrap}
O(\mathbf{x}) = \mathit{\mu}(\mathbf{x})+\varepsilon (\mathbf{x})
\end{equation}
which provides an inexpensive evaluation of  $O$ for hardware parameter setups that are not actually synthesized. Note that $\epsilon$ denotes the error of a  $\mathcal{GP}$'s with a mean function equal to zero and a covariance operator equal to $\mathbf{K}$. 

\begin{figure}[b]
\centering
\includegraphics[width=0.95\linewidth]{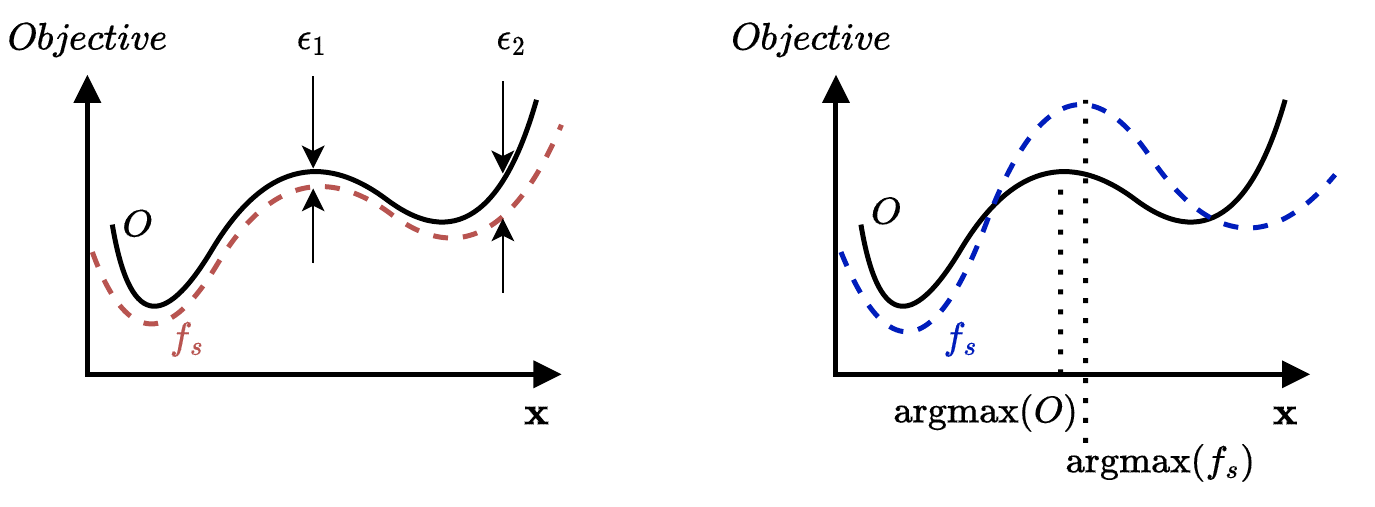}\\
~~~(a)\hspace{4cm}(b)
\caption{(a) Performance modeling (regression) setting where the accuracy of the model is evaluated with the Mean Squared Error (MSE) and (b) Design space exploration setting where the algorithm tries to find the parameters that correspond to the design with maximum performance. }
\label{fig:compare-dse-reg} 
\end{figure} 

\section{Design Space Exploration vs Performance Modeling Settings}\label{sec:dse-vs-reg}
Hardware design space exploration and hardware performance modeling are two related but distinct tasks in hardware design.
Our proposed statistical modeling methodology provides a unified framework for both applications that have been crucial for hardware designers historically. 
In the \textit{Performance Modeling} application, the goal is to accurately \textit{predict} the hardware performance (or value of the objective function $O$) for a specific hardware design parameter setting.  Thus, the performance modeling application is essentially a regression problem and the accuracy of the model can be measured with \textit{Mean Squared Error (MSE)} between the surrogate function $f_s$ and the actual objective function $O$ over $N$ parameter sets $\mathbf{x}_{1:N} = \{\mathbf{x}_1, \mathbf{x}_2, \dots, \mathbf{x}_N\}$

\begin{equation}
\label{eq:mse}
\text{MSE} 
 = \frac{1}{N}\sum_{i=1}^N\left( O(\mathbf{x}_i) - f_s(\mathbf{x}_i) \right)^2 
 = \frac{1}{N}\sum_{i=1}^N\varepsilon_i^2.
\end{equation}
Fig.~\ref{fig:compare-dse-reg} (a) demonstrates the performance modeling setup. The black curve is the actual objective function, while the dotted red curve is the regression model. Note that in our experiments, we report $\sqrt{\text{MSE}}/\mu_{f_s(\mathbf{x}_i)}$ to normalize MSE with respect to the mean of the surrogate function $\mu_{f_s(\mathbf{x}_i)}$.

On the other hand, in the \textit{Design Space Exploration (DSE)} application, the goal is to find the hardware design parameter set that maximizes the objective function. Moreover, in the case of multi-objective optimization, DSE aims to find the dominant or Pareto optimal solutions. Fig.~\ref{fig:compare-dse-reg} (b) shows the DSE setup. The model is accurate if 

\begin{equation}
\label{eq:dse-setup}
{{\arg\max(f_s) = \arg\max}}(O), 
\end{equation}

meaning that the surrogate function $f_s$ can accurately predict the parameter set that maximizes the objective function. Also, note that here \textit{argmax} is used for the objectives that need to be maximized, such as throughput. Likewise, the goal for the objective functions that are to be minimized, e.g. power consumption, is the same, except that $\arg\max$ should be replaced by $\arg\min$. 

\begin{table*}[!b]
\centering
\renewcommand\thetable{1}
\caption{Design space exploration using Bayesian Active Learning and Comparison with previous works.}
\label{table:compare}
\setlength{\tabcolsep}{3pt}
\begin{tabular}{m{95pt}>{\centering}m{60pt}>{\centering}m{60pt}>{\centering\arraybackslash}m{60pt}>{\centering\arraybackslash}m{60pt}>{\centering\arraybackslash}m{60pt}>{\centering\arraybackslash}m{60pt}}
\hline
\textbf{Benchmark Function} & \textbf{\small{Random Search \cite{bergstra2012random} Average  Latency(ms)}} &{\textbf{\small{LASSO Model (ms) \cite{gautier2016spector}}}}& {\textbf{\small{Hybrid GWO-BO \cite{ghaffari2021efficient} Average Latency (ms)}}}& {\textbf{\small{This work Latency (ms)}}} \\
\hline
\hline
\textbf{Breadth-First Search Dense} & 4.372  &  4.3825 & {4.329} & \textbf{4.2804} \\
\textbf{Breadth-First Search Sparse} & 13.371 & 13.2857 & {12.5151} & \textbf{12.5151} \\
\textbf{Discrete Cosine Transform } & 2.904  & 2.9522 & {2.5891} & \textbf{2.1916} \\
\textbf{FIR} & 1.8$\times 10^{-3}$  &  2.4$\times 10^{-3}$  & {1.6${\times 10^{-3}}$} & \textbf{7.11$\mathbf{\times 10^{-4}}$}\\
\textbf{Histogram} & 1.6632 & 1.5715 & {1.5652} &  \textbf{1.5481}\\
\textbf{Mergesort} & 1.9955 & 2.0441 & {1.9386} &  \textbf{1.9269} \\
\textbf{Matrix  Multiplication} & 40.9993 & 40.6412 & {36.8913} &  \textbf{34.198} \\
\textbf{Normal Estimation} & 6.5449 & 6.4763 & {6.4315} &  \textbf{6.2779}\\
\textbf{Sobel Filter} & 1.8448 & 1.9331 & {1.7434} & \textbf{1.719} \\
\textbf{Sparse Matrix Vector Multiplication} & 0.1729 & 0.1798 & {0.1671} &  \textbf{0.1661} \\
\hline

\end{tabular}

\end{table*}

\section{Results and discussions}\label{sec:results}
Here, we discuss the performance of our proposed method and compare it with previous works. We show that our proposed method not only outperforms other algorithms  for design space exploration applications, but it also provides a viable performance prediction in the regression setting.
\subsection{Design Space Exploration (DSE) Setting}
We study the performance of our algorithm for DSE setting for two applications. In Section~\ref{sec:multi-objective-pareto}, we study the multi-objective optimization of OpenCL kernels on FPGA targets, and in  Section~\ref{sec:dse-tl} we show the effect of transfer learning on design space exploration of the CPU micro-architecture design.

\subsubsection{Case Study I: Multi-Objective Optimization of OpenCL kernels on FPGA targets}\label{sec:multi-objective-pareto}

As studied previously in (\cite{gautier2016spector,ghaffari2021efficient}), design space exploration of the OpenCL kernels on FPGA targets is a challenging task that involves optimizing multiple competing performance objectives such as throughput and area utilization  using various design parameters. 
These design parameters include, but are not limited to, the number of work-groups, the number of compute units, SIMD size, unrolling, etc. Furthermore, such kernels must be optimized for both area and throughput. Having two objectives, i.e., logic utilization and throughput, makes the problem a multi-objective problem where the \textit{Pareto Frontier} must be found by the proposed algorithm. To do so, we create a  Bayesian active learning multi-model for throughput and area according to Algorithm~\ref{alg:al_alg}. 

The idea behind using our proposed method is to iteratively select new design parameters and configure the application based on them in order to find the optimal set of parameters that achieve the required performance. As shown in Algorithm~\ref{alg:al_alg}, the Bayesian models of area utilization and latency are trained on a set of initial design parameters and their corresponding performance metrics using Gaussian processes. Then, these models are used to model the performance of other design configurations and also as a guide for the selection of new parameters to be evaluated by the Bayesian model. By using active learning, the optimization process proceeds to update the Bayesian model based on the new samples evaluated by the model and it continues until satisfactory results are achieved for the performance metrics. For this experiment, we retain the five candidate solutions defined by a set of design parameters that give the best score according to the surrogate function. More generically, there could be more than one objective function, in which case each objective function would point to the five best designs. Finally, the associated \textit{parameter sets} producing the best performance in the surrogate models are used to update the Bayesian models in each active learning round. 


\begin{figure}[h]
\vspace{-2.5cm}
\centering
\includegraphics[width=1\linewidth]{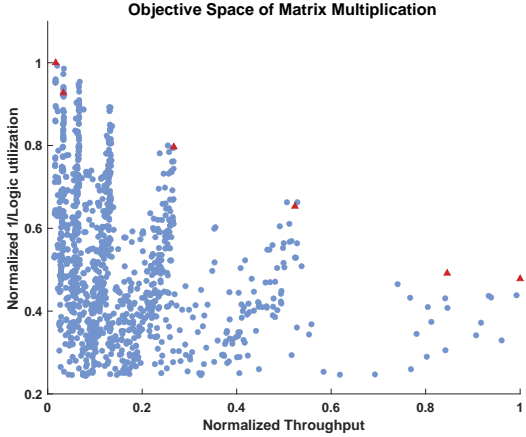}
\vspace{-2.7cm}
\caption{Objective space of the Matrix Multiplication OpenCL kernel is demonstrated by the blue dots. The Pareto frontier estimated by our proposed multi-model active learning method is depicted with red triangles.}
\label{fig:mm-pareto} 
\end{figure}

Fig. \ref{fig:mm-pareto} shows the objective space of the matrix multiplication OpenCL kernel\footnote{For a complete list of matrix multiplication HLS design parameters, refer to \textbf{Appendix A}.}. The red triangles show the set of dominant solutions, also known as Pareto frontier \cite{ngatchou2005pareto} that is estimated by our proposed method. The estimated Pareto frontier resembles the actual Pareto frontier in \cite{gautier2016spector}.

Table \ref{table:compare} shows the performance of the proposed method when optimizing the OpenCL kernels for latency. Our proposed method outperforms all the previously reported methods to find the optimum parameters for achieving the best latency results in various OpenCL benchmarks.

\subsubsection{Case Study II: Design Space Exploration of Micro-Architectures using Transfer Learning}\label{sec:dse-tl}

Design space exploration of micro-architectures is one of the most prominent problems of digital hardware design. Similar to other design space exploration problems, the number of possible designs grows exponentially with the number of parameters (See Appendix A for the list of design parameters). As such, finding the optimal design parameters that meet the performance requirement is a challenging and time-consuming task that requires significant computational and synthesis resources. Thus, there is a need to find performance models with a minimum number of queries for the synthesis tool. In order to show that our proposed algorithm performs efficiently in this setup, similar to \cite{lee2006accurate}, we consider the SPEC2k \cite{Spec2k} benchmarks to evaluate our proposed methodology. More specifically, we use the ammp, applu, equake, gcc, gzip, mesa, twolf and jbb benchmarks.  The SPEC2k dataset exposes twelve parameters that can be tuned to optimize performance. These parameters are included but not limited to the cache size, the memory latency, the fixed-point and floating-point latency, etc. \cite{lee2006accurate}. Moreover, considering the number of parameters, the design  space for each benchmark is about \textit{one Billion} design points, which makes it impossible to do design space exploration using exhaustive search methods such as brute force.

\begin{figure}[t]
\vspace{-2.5cm}
\centering
\includegraphics[width=0.95\linewidth]{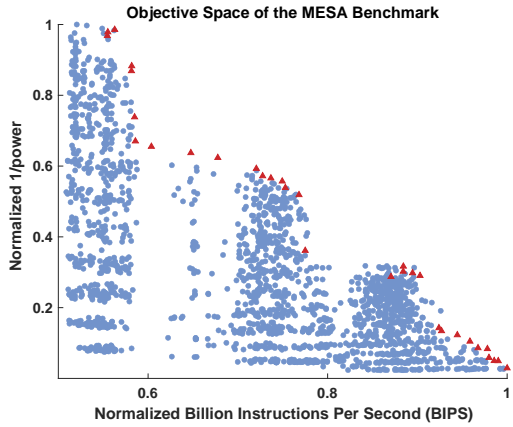}
\vspace{-2.7cm}
\caption{Selected samples of the objective space for the SPEC2k 3-D graphics benchmark (mesa) demonstrated by the blue dots. The estimated Pareto frontier obtained with our proposed multi-model active learning method is depicted with red triangles.}
\label{fig:mesa-pareto} 
\end{figure}

Similar to Section \ref{sec:multi-objective-pareto}, we can consider the design space exploration of the micro-architectures as a multi-objective problem and illustrate the dominant solutions using a Pareto frontier. Fig.~\ref{fig:mesa-pareto} shows a selected group of samples for objective space for SPEC2k 3-D graphics benchmark (mesa). In this figure, the Pareto frontier shows the trade-off between the Billion Instructions per Second (BIPS) rating and the normalized power, confirming that a faster microprocessor needs to consume more power. Also note that the red dots in Fig.~\ref{fig:mesa-pareto} are determined by our proposed Bayesian multi-model active learning of power consumption and instructions per second (BIPS) rating created using 600 samples. As discussed in Section \ref{sec:regression}, our proposed method considerably reduces the number of samples needed to perform regression analysis compared to the method proposed by \cite{lee2006accurate}. 

Now we show that Bayesian Transfer Learning, as introduced in Section \ref{sec:methodTL}, can further reduce the number of samples required for design space exploration. We use transfer learning as a technique to leverage knowledge gained from training Bayesian models on a source task to improve the modeling process of a target task. Let us consider the SPEC2k Java Business Benchmark (jbb) as the source task $\mathcal{T}_{source}$ and show that the information of its Gaussian surrogate function can be used to reduce the number of samples needed to perform design space exploration for other benchmark tasks. Our approach for transfer learning in this context is to use a Bayesian model that is trained using SPEC2k Java Business Benchmark (jbb) and use its Gaussian process information to affect the training of other target tasks as explained in Section~\ref{sec:methodTL}. By doing so, the number of design evaluations is reduced, which leads to faster and more efficient design space exploration.

Note that the performance results of a micro-architecture on various benchmark tasks are strongly correlated. This means that we expect a powerful microprocessor to perform better in all tasks. This correlation is a crucial characteristic of design space exploration that can be exploited by inductive transfer learning. Table \ref{table:corr} shows this phenomenon for our selected SPEC2k benchmark tasks with the SPEC2k jbb benchmark. Note that small \textit{pvalues} in Table \ref{table:corr} shows that the correlation results are statistically significant. In other words, it confirms the hypothesis that there is a strong correlation between the metric performance of the considered tasks.

\begin{table}[!t]
\centering
\renewcommand\thetable{2}
\caption{Correlation of the SPEC2k benchmarks with SPEC2k Java Business Benchmark (jbb).}
\label{table:corr}
\setlength{\tabcolsep}{3pt}
\begin{tabular}{lcc}
\hline
\textbf{Benchmark Task (BIPS)} & \textbf{Correlation ($\rho$) }& \textbf{pvalue} \\
\hline
\hline
\textbf{ammp} & 0.6395  &  3.25$\times 10^{-230}$  \\
\textbf{applu} & 0.7523  &  0  \\
\textbf{equake} & 0.9100  &  0  \\
\textbf{gcc} & 0.9397  &  0  \\
\textbf{gzip} &  0.8469  &  0  \\
\textbf{mesa} & 0.8720  &  0  \\
\textbf{twolf} & 0.8920  &  0  \\
\hline
\end{tabular}
\end{table} 

Table \ref{table:dse-spec2k} shows the experimental results for design space exploration using transfer learning for different target tasks where the source task is the SPEC2k jbb benchmark. The numbers reported in the table indicate a considerably faster convergence to the desired BIPS. 
For this experiment, $\lambda_2=0$ and $\lambda_1$, in eq \eqref{eq:tfl}, are both set to $0.5$ at the beginning of the iterations and linearly decreased to zero through the 
course of iterations. Also note that each iteration is a query to the synthesis environment, each of which consists of five samples, to create the Gaussian processes. 

\begin{table}[!t]
\centering
\renewcommand\thetable{3}
\caption{Design space exploration of the SPEC2k benchmarks for the Billions of Instructions Per Seconds (BIPS) performance.}
\label{table:dse-spec2k}
\setlength{\tabcolsep}{3pt}
\begin{tabular}{lccc}
\hline
                         &               & ~\textbf{Number of  Samples} ~& \textbf{Number of Samples} \\
 \textbf{Benchmark Task} & ~\textbf{BIPS}~ & \textbf{without} & \textbf{with}  \\
                          &              & \textbf{Transfer Learning} & \textbf{Transfer Learning}  \\
\hline
\hline
\textbf{ammp} & 1.302 & 35  & 25   \\
\textbf{applu} & 1.060 & 115  & 100 \\
\textbf{equake} & 1.027& 15 & 15 \\
\textbf{gcc} & 1.009 & 20  & 15\\
\textbf{gzip} & 1.005 & 25  & 15  \\
\textbf{mesa} & 2.012 & 25 & 20 \\
\textbf{twolf} & 1.138 & 20 & 15 \\
%
%
%
\hline
\end{tabular}
\end{table} 

\begin{table*}[!t]
\centering
\renewcommand\thetable{4}
\caption{Comparison of the Predictive power of regression models with or without using Gaussian regression bootstrapping on the SPEC2k AMMP benchmark.}
\label{table:regression-mse}
\setlength{\tabcolsep}{3pt}
\begin{tabular}{lcc}
\hline

                         &                             & \textbf{Regression using} \\
                         & \textbf{Original}                    &  \textbf{Gaussian Bootstrapping}\\
                         &          \textbf{Regression Setting} &  \textbf{Simulated Data}\\
                         &          $\sqrt{\text{MSE}}/\mu$ &  $\sqrt{\text{MSE}}/\mu$\\
\hline
\hline
\textbf{Linear Regression Instructions Per Second} & 0.056   & 0.061   \\
\textbf{Linear Regression Power Consumption (mW)} & 0.45 &  0.46   \\
\textbf{Random Forest Instructions Per Second} & 0.017  &  0.055  \\
\textbf{Random Forest Power Consumption (mW)} & 0.050  & 0.14   \\
\hline
\end{tabular}
\end{table*} 

\begin{figure*}[!t]
\vspace{1cm}
\centering
\includegraphics[width=0.4\linewidth]{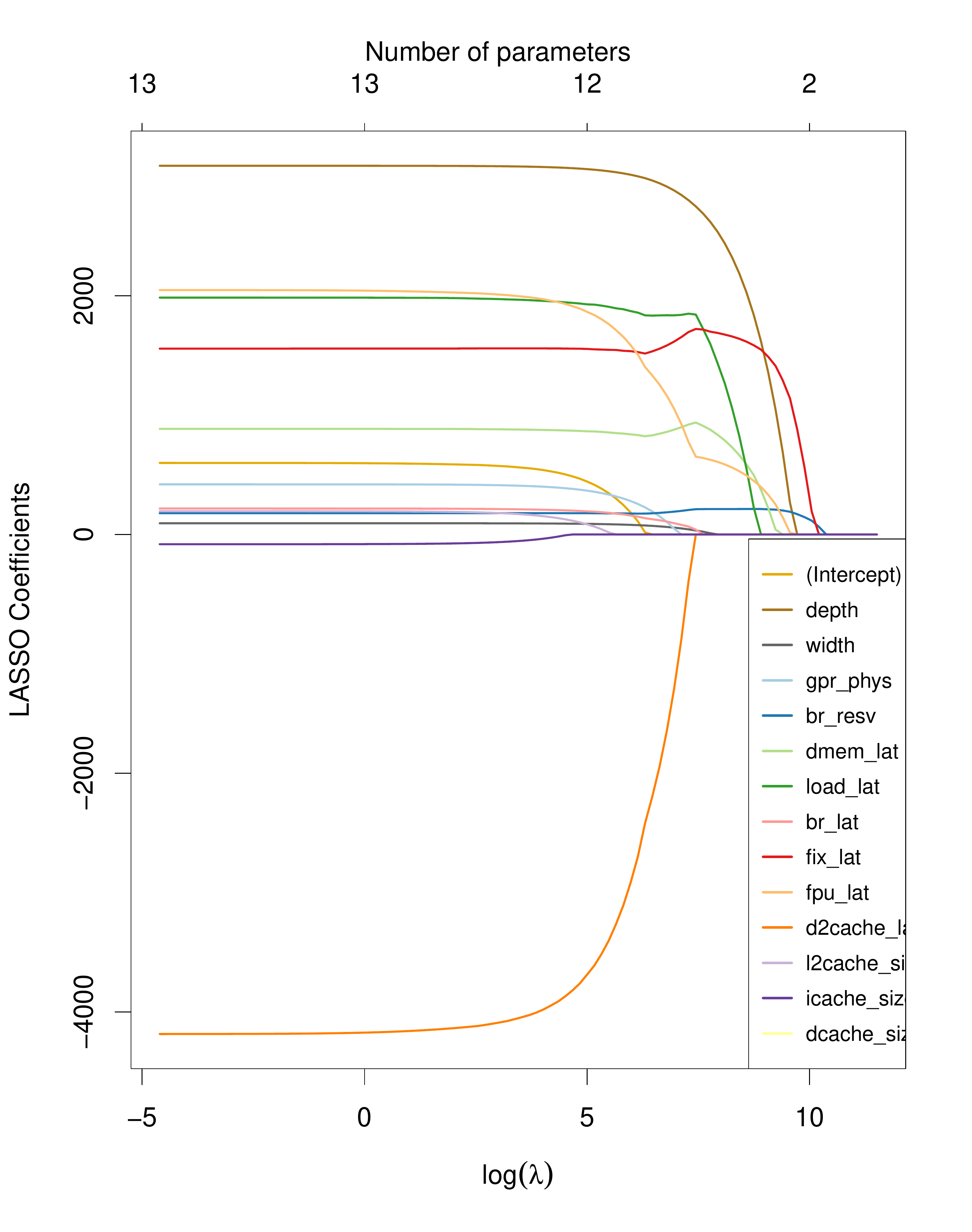}
\hspace{1.5cm}
\includegraphics[width=0.4\linewidth]{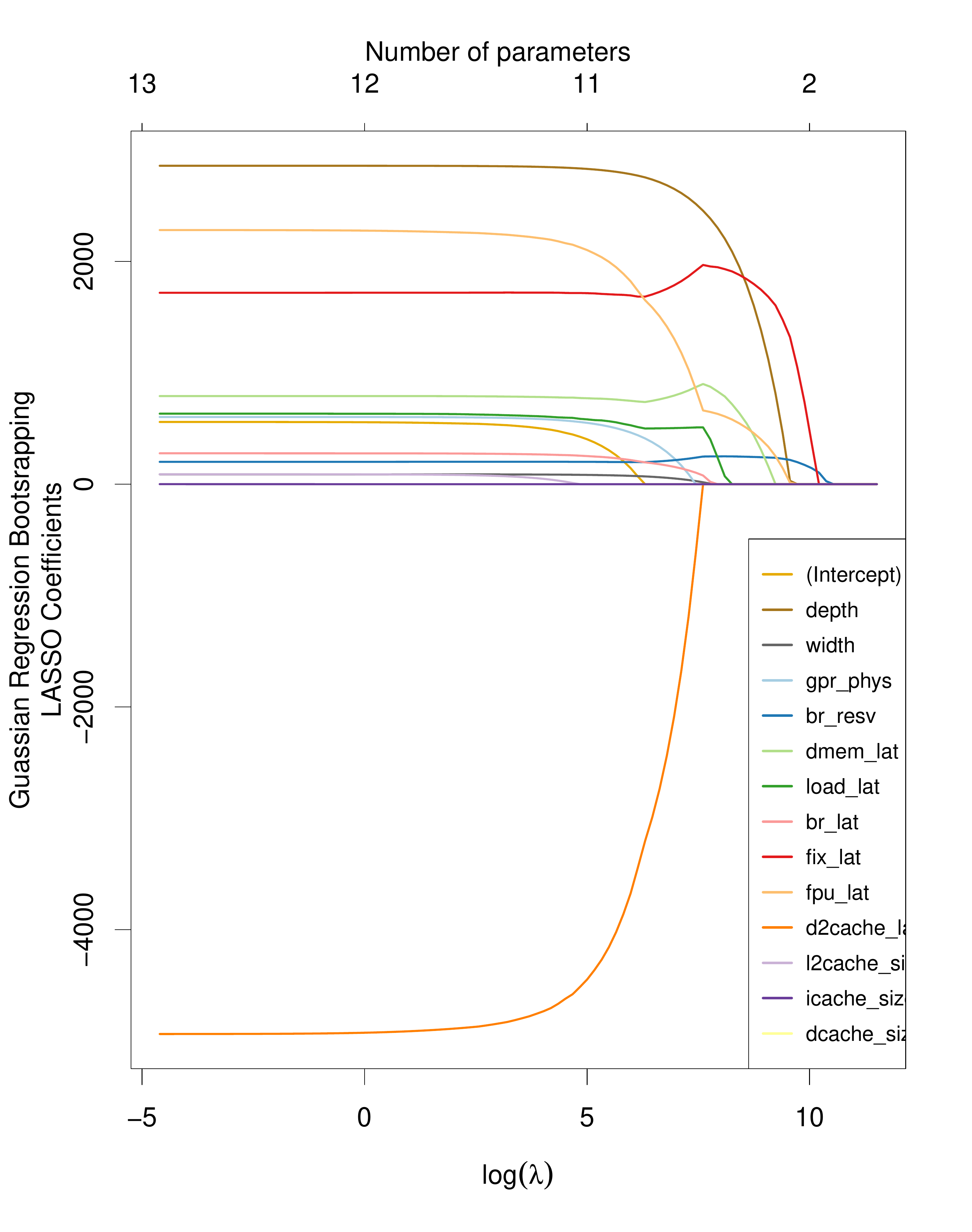}\\
\hspace{2cm}(a)\hspace{8cm} (b)\hspace{1cm}
\caption{ (a) Original LASSO model with a shrinkage factor $\lambda$ for the coefficients of the power consumption model produced using 4000 samples (b) LASSO shrinkage factor $\lambda$ effect on the coefficients of the power consumption model that is created using 700 samples and Gaussian regression bootstrapping. The comparison shows the extent to which Gaussian regression bootstrapping is efficient in producing simulated data samples to create statistical hardware performance prediction models. }
\label{fig:lasso} 
\end{figure*}

\subsection{Gaussian Regression Bootstrapping and Performance Prediction Models}\label{sec:regression}

Bootstrap is a resampling technique in statistics that allows one to create simulated data from a given set of samples. In our proposed method, when the Bayesian models are created using Gaussian processes, it is possible to use such probabilistic methods to create simulated data that closely resemble the behavior of the hardware. Gaussian regression bootstrapping can be used in various contexts such as bioinformatics \cite{GPRbootstrap}. 

The Gaussian regression bootstrapping involves fitting Gaussian process regression to a set of performance data and using this model to resample simulated data to create new statistical models. This allows quantifying the uncertainty in the data and improves the accuracy of the statistical modeling for performance prediction models. The simulated data that is generated by Gaussian regression bootstrapping can be used in a variety of statistical analyses to predict the behavior of the hardware. For instance, we show that Gaussian regression bootstrapping can generate robust simulated data that can be efficiently used in various regression analyses such as linear regression, LASSO, and Random Forest regression.

Here we explore the effectiveness of our proposed active learning multi-model to predict the performance of hardware benchmarks. As mentioned previously in Section \ref{sec:dse-vs-reg}, the regression setting is substantially different from design space exploration and requires more samples from the actual hardware to provide acceptable models. For instance, in the case of SPEC2k benchmarks, the authors of \cite{lee2006accurate} show that 2000 samples are enough to create performance regression models. However, using our proposed methodology, it is possible to reduce this number to 700 samples using Gaussian regression bootstrapping which is 65\% sample reduction.

\subsubsection{Linear Regression}
Linear regression is the most basic tool for the statistical modeling of hardware. The underlying assumption, of course, is that the performance of the hardware under study is a linear function of the design parameters. Linear regression can be used to develop performance models of hardware based on tuning various design parameters such as cache size, floating point delay, etc.  Table \ref{table:regression-mse} shows the comparison of $\sqrt{\text{MSE}}/\mu$ for linear regression using 2000 samples suggested in \cite{lee2006accurate} and 700 samples using our proposed Gaussian regression bootstrapping method.

\subsubsection{Random Forest}

Random forest \cite{liaw2002classificationRandomF} is a tree-based regression method that has been used previously for design space exploration of hardware \cite{liu2013learning}. Given the non-linear nature of hardware design, non-linear regression methods, such as Random Forest, might behave better in such scenarios. In random forest regression, multiple decision trees are created, and each tree chooses a random subset of design parameters from the training data to fit a statistical model. The main advantage of Random Forest regression over linear regression is that it can model the complex and nonlinear relation between the parameters and the desired performance objective. Table \ref{table:regression-mse} reports that the performance of the random forest regression based on simulated data using  the Gaussian regression bootstrapping is close to the original Random Forest regression. This means that the random forest regression on the simulated data and actual data produces similar results.

Note that the $\sqrt{\text{MSE}}/\mu$  values in Table  \ref{table:regression-mse} are reported for the case when the covariance kernel function is squared exponential. We show in \textbf{Appendix B} that using other types of covariance kernel (such as Matern covariance kernels) for this specific application has a negligible effect on the reported results.

\subsubsection{LASSO}

The Least Absolute Shrinkage and Selection Operator (LASSO) \cite{tibshirani1996regressionLASSO} is a  regularization and variable selection method that is used in regression analysis. 
LASSO is also used in \cite{gautier2016spector} for performance modeling of hardware. Furthermore, LASSO can be used in hardware performance prediction models to identify the most important design parameters by leveraging the variable selection property of the LASSO. The variable selection of the LASSO is particularly important for hardware performance prediction models when working with large datasets with many design parameters since the inclusion of irrelevant parameters in the performance model can lead to over-fitting the hardware performance model. Fig. \ref{fig:lasso}.a  shows that as the LASSO shrinkage factor $\lambda$ increases, the design parameter coefficients shrink to zero. The first coefficient that collapses to zero identifies the least significant parameter in the design space and also the last parameter that collapses to zero is the most important parameter in the design space. The top axis in this figure shows the number of variables that are not zero at the given value of $\lambda$. Fig. \ref{fig:lasso}.b depicts 
the behavior of the design parameter coefficients for the regression model created from simulated data using Gaussian regression bootstrapping. Comparing Fig. \ref{fig:lasso}.a and Fig. \ref{fig:lasso}.b reveals that the behavior of the Bayesian model created using simulated data  and the original LASSO model are similar for different values of the LASSO shrinkage factor, $\lambda$. For instance, in both LASSO models, \textit{`Fixed point latency'} and \textit{`Number of reservation stations'} are the last design parameter coefficients that are shrunk to zero. Also, note a complete list of design parameters is reported in \textbf{Appendix A}. This shows that Gaussian regression bootstrapping is efficient in producing simulated data samples to create a reliable statistical model for hardware performance prediction.

\section{Conclusion}
We proposed a multi-model Bayesian active learning framework to statistically model digital hardware performance. Our proposed model uses Gaussian processes to characterize the performance of hardware designs. We also use active learning to iteratively guide the sampling process in order to reduce the number of samples needed to create the statistical model. Furthermore, we used Bayesian transfer learning in order to incorporate  the prior design knowledge acquired in a source application to better model a target application with fewer data samples in the design space exploration task. We further propose using the Gaussian regression bootstrapping method to reduce the number of samples required for the performance prediction task. To show the effectiveness of our proposed method, we performed design space exploration and performance prediction for various hardware setups, such as micro-architecture design and OpenCL kernels. For OpenCL kernels on FPGA targets, our approach was able to identify the minimum latency in all benchmarks as well as predict the correct Pareto frontier. Our experiments show that our approach can detect the optimal design parameters for processors' micro-architectures with less than 50 samples in most cases. Our experiments also show that the number of samples required to create performance models significantly reduces, by 65\%, while maintaining the model's predictive power. We have also performed commonplace regression analysis, such as linear regression, LASSO, and random forest regressions,  using simulated data generated by Gaussian regression bootstrapping and demonstrated that our proposed statistical hardware models closely follow the actual hardware behavior in various scenarios.

\section{Acknowledgments}
\noindent The authors would like to thank Dr. Vahid Partovi Nia whose constructive comments improved the presentation and exposition of this manuscript.

\bibliographystyle{IEEEtran}
\bibliography{biblio.bib}

\begin{table*}[!b]
\centering
\renewcommand\thetable{7}
\caption{Comparison of Matern kernels vs squared exponential kernel on Gaussian regression bootstrapping on the SPEC2k AMMP benchmark.}
\label{table:regression-mse-matern}
\setlength{\tabcolsep}{3pt}
\begin{tabular}{lcccc}
\hline
\textbf{ Covariance Kernel}&                             & Squared Exponential & Matern ($\nu=3/2$)~~~~ & ~~~~(Matern $\nu=5/2$)\\
\hline
                         & Original                    &  Gaussian Bootstrapping&  Gaussian Bootstrapping&  Gaussian Bootstrapping\\
                         &          $\sqrt{\text{MSE}}/\mu$ &  $\sqrt{\text{MSE}}/\mu$&  $\sqrt{\text{MSE}}/\mu$&  $\sqrt{\text{MSE}}/\mu$\\
\hline
\hline
\textbf{Linear Regression Instructions Per Second} & 0.056   & 0.061  & 0.061 & 0.062 \\
\textbf{Linear Regression Power Consumption (mW)} & 0.45 &  0.46  & 0.46 & 0.46 \\
\textbf{Random Forest Instructions Per Second} & 0.017  &  0.055 & 0.052 & 0.056 \\
\textbf{Random Forest Power Consumption (mW)} & 0.050  & 0.14  & 0.13& 0.13\\
\hline
\end{tabular}
\end{table*}

\begin{table}[!h]
\centering
\renewcommand\thetable{5}
\caption{OpenCL Matrix multiplication design parameters.}
\label{table:opencl_params}
\setlength{\tabcolsep}{3pt}
\begin{tabular}{lc}
\hline
\#&\textbf{Parameter}\\
\hline
\hline
1&number of blocks\\
2&Sub-dimension x\\
3&Sub-dimension y\\
4&Manual SIMD x\\
5&Manual SIMD y\\
6&SIMD\\
7&Number of compute units\\
8&Enable Unroll\\
9&Unroll factor\\
\hline
\end{tabular}
\end{table} 

\begin{table}[!h]
\centering
\renewcommand\thetable{6}
\caption{Spec2k design parameters.}
\label{table:spec_params}
\setlength{\tabcolsep}{3pt}
\begin{tabular}{lcc}
\hline
\#&\textbf{Parameter}\\
\hline
\hline
1&  FO4 depth\\
2&  Load/store queue width\\
3&  Number of physical registers\\
4&  Number of reservation stations\\
5&  i-L1 cache\\
6&  d-L1 cache\\
7&  L2 cache\\
8&  Control latency\\
9&  Floating point latency\\
10&  Fixed point latency\\
11&  Load/store latency\\
12&  Memory latency\\

\hline
\end{tabular}
\end{table} 

\section{\textbf{Appendix A:} Hardware design parameters}

The HLS design parameters of OpenCL matrix multiplication kernel are listed in Table \ref{table:opencl_params}. Additionally, the design parameters of SPEC2k benchmark are reported in Table \ref{table:spec_params}.

\section{\textbf{Appendix B:} Bayesian Optimization using Matern covariance kernel}
In Bayesian optimization, choosing a proper covariance kernel is essential for accuracy of the algorithm. Matern kernels are a class of stationary covariance function that depends on two hyper-parameters known as the length scale $l$ and the smoothness $\nu$. By choosing a suitable value for $\nu$ the Bayesian model can capture the smoothness and roughness of the objective function. The Matern class of covariance kernels is defined as
 \begin{align}
\label{eq:cov_matrix_matern}
&\mathbf{K}_{ij}=\mathit{k_{\text{Matern}}}({x}_i, {x}_j) \\ \nonumber
& =\frac{1}{2^{\nu-1}\Gamma(\nu)} \left ( \frac{\sqrt{2\nu}|{x}_i- {x}_j|}{l} \right)^\nu {H}_v\left ( \frac{\sqrt{2\nu}|{x}_i- {x}_j|}{l} \right),
\end{align}
where $\Gamma$ is the Gamma function and $H_v$ is modified Bessel function of the second kind of order $\nu$. A common choice for $\nu$ are $\frac{3}{2}, \frac{5}{2}$ and also when $\nu \to \infty$ the Matern kernels becomes the squared exponential kernel \cite{genton2001classes}.

We repeated the experiments in Table \ref{table:regression-mse} using Matern covariance kernels with $\nu \in \left\{\frac{3}{2}, \frac{5}{2}\right\}$ and reported the results in Table \ref{table:regression-mse-matern}. The reported $\sqrt{\text{MSE}}/\mu$ in Table \ref{table:regression-mse-matern} shows that our method exhibits a robust behavior in terms of choosing various covariance kernels for modeling SPEC2k ammp benchmark. Note that the choice of covariance kernel depends on the nature of the data and can vary in different applications.

\end{document}